\documentclass[apj]{emulateapj}
\def\mathbi#1{\textbf{\em #1}}
\let\oldhat\hat
\renewcommand{\vec}[1]{\mathbi{#1}}
\renewcommand{\hat}[1]{\oldhat{\mathbf{#1}}}

\usepackage{natbib}
\usepackage{natbib}
\usepackage{graphicx}
\usepackage{subfigure}
\usepackage{epsf}
\usepackage{amsmath}
\usepackage{gensymb}

\newcommand\vlsr{v_{\rm LSR}}
\newcommand\vdev{v_{\rm DEV}}
\newcommand\Msun{\; {\rm M}_{\odot}}
\newcommand\kms{\; {\rm km}\;{\rm s}^{-1}}
\newcommand\nhi{n_{\rm HI}}
\newcommand\Nhi{\rm N_{\rm HI}}
\newcommand\Msunyr{\; {\rm M}_{\odot} {\rm yr}^{-1}}
\newcommand\vmin{v_{\rm min}}
\newcommand\vmax{v_{\rm max}}
\newcommand\Mdot{\dot{M}}

\begin{document}

\title{The Circumgalactic Medium of the Milky Way is Half Hidden}

\author{Y. Zheng\altaffilmark{1}, M. E. Putman\altaffilmark{1}, J. E. G. Peek\altaffilmark{1,2}, M. R. Joung\altaffilmark{1}}
\altaffiltext{1}{Department of Astronomy, Columbia University, New York, NY 10027, USA}
\altaffiltext{2}{Space Telescope Science Institute, 3700 San Martin Dr, Baltimore, MD 21218, USA}

\begin{abstract}
We assess the fraction of the Milky Way's circumgalactic medium (CGM) eluding detection due to its velocity being similar to gas in the disk. This is achieved using synthetic observations of the CGM in a simulated MW-mass galaxy that shows similar CGM kinematics to the MW and external L$\sim$L$_*$ galaxies. As viewed by a mock observer at a location similar to the Sun, only 50$\%$ (by mass) of the gas moves at high velocity ($|\vlsr|\geq$100$\kms$ or $|\vdev|\geq$50$\kms$) in the simulated CGM and would be observable. The low velocity gas is thermodynamically similar to the high velocity gas, indicating the 50\% observable fraction is applicable to spectral lines from the radio to the ultraviolet. We apply the observable mass fraction (50$\%$) to current estimates of the MW's CGM, and find a corrected total mass of 2.8$\times$10$^{8}\Msun$ for gas below 10$^6$K within $\sim15$ kpc (this excludes the Magellanic System). This is less than the total mass of the CGM extending out to $\sim$150 kpc in other L$\sim$L$_*$ galaxies. However, we find similar \ion{O}{6} column densities when the discrepancy in path length between the MW and external galaxies is considered. The coherent spatial and kinematic distribution of low velocity gas in the simulated CGM suggests that current \ion{H}{1} observations of the MW's CGM may miss large low velocity \ion{H}{1} complexes. In addition, current mass estimates of the MW's CGM based on high-velocity line observations with distance constraints may miss a non-negligible fraction of gas in the outer halo which can be obscured if it moves at a velocity similar to the gas in the lower halo.

\end{abstract}


\section{INTRODUCTION}
\label{sec1}
Studies of the circumgalactic medium (CGM) help us to understnad the origin of a galaxy's star formation fuel and the impact of galactic feedback. Assessing the amount and state of the baryons in galaxy halos is a key active area in galaxy formation research. Due to its proximity, the Milky Way (MW)'s CGM has been studied in the most detail, and this has helped us interpret the nature of the multiphase medium and the potential interactions with the galactic disk.  

To calculate the mass of the multiphase CGM in the MW, numerous spectroscopic observations towards high-velocity line features have been conducted. High velocity clouds (HVCs), mapped in \ion{H}{1} surveys via the 21 cm hyperfine transition line, trace the coldest component of the Galactic halo. HVCs have been found to largely reside within 15 kpc of the disk with the exception of the complexes associated with the Magellanic System which are at least at $\sim$55 kpc (\citealt{Wakker97}; \citealt{Putman12}, hereafter PPJ12). The \ion{H}{1} mass of the non-Magellanic System HVCs is $\sim$3$\times$10$^7\Msun$ (PPJ12), and thus is a very small fraction of the Galaxy's baryons. Observations of the ion absorption lines in the spectra of background quasars show an order of magnitude more mass is associated with the warmer ionized gas (T$\sim$10$^{4-6}$ K) in the Galactic halo (\citealt{Sembach03}, \citealt{Shull09}, PPJ12). X-ray observations of hot gas (T$>$10$^6$ K) in emission and absorption (\ion{O}{7}, \ion{O}{8}) do not have strong constraints from spectral line observations, but currently indicate a mass of $\sim$10$^{10}\Msun$ if extended out to 250 kpc \citep{Anderson10, Miller13, Miller14}.

The MW's CGM can be compared to that of other L$\sim$L$_*$ galaxies now with the increase in absorption line probes made available with the Cosmic Origins Spectrograph (COS). Discrepancies in both mass and column density have been found. At first glance, the total mass of the MW's CGM with T$\leq$10$^6$ K appears to be 1-2 orders of magnitude lower than the CGM mass of other L$\sim$L$_*$ spirals ($\sim$10$^{10}\Msun$; \citealt{Stocke13}, \citealt{Werk14}). As for the column density, \cite{Tumlinson11} found a typical value of ${\rm log\ N_{OVI}}$=14.5 in the CGM of L$\sim$L$_*$ galaxies with specific star formation rate (sSFR)$>$10$^{-11}$ yr$^{-1}$ in the COS-Halos sample, while the MW's CGM instead has ${\rm log\ \big \langle N_{OVI}\big \rangle}$=14.0 \citep{Sembach03} with sSFR$\sim$2-6$\times$10$^{-11}$ yr$^{-1}$ (SFR$\sim$1-4$\Msun$ yr$^{-1}$, \citealt{Robitaille10, Diehl06}; ${\rm log\ M_*}$ $\sim$10.8, \citealt{McMillan11}). These differences between the MW and external L$\sim$L$_*$ galaxies raise several questions. Is the MW's CGM an outlier with much less mass than other L$\sim$L$_*$ spirals? Or are our observations of the MW's CGM still incomplete? 

Observations of the MW's CGM specifically suffer from obscuration by the dense foreground disk gas that can be moving at a similar radial velocity as the gas in the MW's CGM. The disk gas causes studies of the MW's CGM to apply velocity cutoffs with commonly adopted values of $|\vlsr|<90-100\kms$ (\citealt{Wakker91b}, \citealt{Sembach03}, \citealt{Lehner12} and PPJ12). It is also common to exclude gas with deviation velocity $|\vdev|<50-75\kms$ which is the amount a cloud's velocity deviates from a simple disk rotation model (\citealt{Wakker91b}, PPJ12). However, estimates of the obscured MW's CGM mass due to these velocity cutoffs are rarely presented since it is usually impossible to distinguish signals of low velocity halo gas from that of disk gas.  

In this work, we conduct synthetic observations of the CGM of a simulated MW-mass galaxy. Observational conditions of the MW's CGM are reproduced so that the influence of foreground disk material on the observed CGM properties can be evaluated. Following the definitions in \cite{Wakker91b}, two velocity cutoffs ($|\vlsr|\geq100\kms$ and $|\vdev|\geq50\kms$) are adopted to define the high velocity gas that is observable in the simulated CGM. We do not perform sophisticated modelling of the gas ionization state, but rather approximate the CGM gas phases by several temperature ranges: cold, T$\leq$10$^4$ K; warm, T$\sim$10$^{4-5}$ K; warm-hot, T$\sim$10$^{5-6}$ K; hot, T$>$10$^6$ K. These loosely correspond to gas traced by atomic hydrogen (\ion{H}{1}), low and intermediate ions (e.g., \ion{Si}{2}, \ion{Si}{3}, \ion{Si}{4}, \ion{C}{2}, \ion{C}{3}, \ion{C}{4}), high ions (e.g., \ion{O}{6}), and X-ray emitting/absorbing ions (e.g., \ion{O}{7}, \ion{O}{8}),  respectively. With this setup, we can understand the CGM distribution in high and low velocity regimes at various phases, which will shed light upon the question of whether the MW's CGM is abnormal or current observations are incomplete. 

This paper is organized as follows: Section \ref{sec2} briefly describes the setup of the numerical simulation and gas kinematics of the simulated CGM; Section \ref{sec3} outlines the methods we use to conduct synthetic observations on the simulated CGM with a view from within the simulated disk. The properties of obscured low velocity gas in the simulated CGM through spectroscopic observations are shown in Section \ref{sec4}, and comparisons with current observations of the MW and other L$\sim$L$_*$ galaxies are discussed in Section \ref{sec5}. We conclude in Section \ref{sec6} with our main findings. 

\section{Simulation of a MW-mass Galaxy}
\label{sec2}

Our investigation of the gas kinematics in the CGM is based on a high-resolution cosmological simulation of a MW-mass galaxy. The simulation was performed by \cite{Joung12} with {\it Enzo}, an Eulerian hydrodynamics adaptive refinement code. A low-resolution run was first performed which identified four MW-mass halos in Local-Group-like volumes; then one of the halos was re-simulated, achieving a mass resolution of $\sim$10$^5\Msun$ and spatial resolution of 136-272 pc physical (or better at all times). The code includes metallicity-dependent cooling, photoionization by UV extragalactic radiation, shielding of UV radiation by \ion{H}{1}, and photoelectric heating of the interstellar medium. Star formation and stellar feedback (but no AGN feedback) are implemented. We refer the reader to the work done by \cite{Joung12} and \cite{Fernandez12} for detailed information on the performance of the simulation, and only report their key results here. 

The total mass of the simulated dark matter halo is 1.4$\times$10$^{12}\Msun$ within its virial radius (250 kpc). Warm-hot and hot ionized gas dominate the mass accretion over large radii with an accretion rate of $\Mdot$=3-5$\Msunyr$. Nearly 70$\%$ of such flow enters the virial radius along filamentary structures. This inflowing gas interacts with existing halo gas, with its cooling rate and heating rate generally balanced over large radii. Only a small fraction of the filamentary flows manage to reach regions close to the disk in the form of cold clouds (\ion{H}{1}). Another source of \ion{H}{1} is the stripped gas from satellites as they move through the simulated CGM. Inflows and the stripped satellite gas contribute comparable amounts of \ion{H}{1} to the halo. Galactic winds also affect the gaseous CGM. Large-scale collimated outflows driven by supernova feedback have been reported in the simulated galaxy (also see Section \ref{sec3}); these winds generally carry very hot metal-enriched gas and move at high velocities extending from 200$\kms$ to 1000$\kms$ (PPJ12).

\section{Synthetic Observations}
\label{sec3}
In this paper, the {\it simulated disk} is defined as a cylindrical region with a size of R$\leq$18 kpc and ${\rm |z|}\leq$2 kpc following the terminology used in \cite{Joung12}. The {\it simulated CGM} defines a spherical volume with a radius of 250 kpc with the central simulated disk manually masked. We use the {\it yt} analysis toolkit \citep{Turk11} to study the properties of the simulated CGM from a view within the simulated disk, and perform analyses on the CGM gas with all the temperature available in the simulation unless otherwise specified. The synthetic observations account for the galactic differential rotation of the simulated disk and the off-center location of the mock observer (Local Standard of Rest; LSR), which is the case for observations of the MW's CGM. 

We construct a mock Galactic coordinate system analogous to that of the MW, and specify observing sightlines using galactic longitude $l$ and galactic latitude $b$. We then build a disk rotation model by synthesizing \ion{H}{1} 21cm emission from the simulated disk in order to determine the velocity ranges allowed by disk rotation, within which low velocity gas in the simulated CGM at similar velocities will be obscured.

\subsection{Constructing the Mock Galactic Coordinates System in the Simulated Galaxy}
\label{sec3.1}
The simulated disk is centered at the point of maximum baryon density, close to the center of mass of the disk. The rotation axis of the simulated disk is aligned with the angular momentum vector of the disk cylinder. In the mid-plane of the simulated disk, we place a mock observer at a distance of R$_0$=8.0 kpc \footnote{We note that R$_0$=8.0 kpc is slightly shorter than the current best estimate of the Galactic distance of the Sun (R$_{0}$=8.34$\pm0.16$ kpc, \citealt{Reid14}). We calculate the observable mass fraction $f_m$ (see Section \ref{sec4.2} for detailed description of $f_m$) with 8.34 kpc and find it only changes by 0.15$\%$, thus the effect of increasing R$_0$ by 0.3 kpc is negligible.} from the disk center. This location is used to set up the LSR frame for the synthetic observations, which is the reference point for the kinematic properties of the simulated CGM unless otherwise specified. In Section \ref{sec4.2.3}, we will additionally examine whether the observed kinematic properties depend on the location of the LSR by relocating the mock observer to seven other locations in the mid-plane at the same galactocentric distance. 

Originating from the LSR, we define \vec{\rm UVW} axes under the right-handed convention: \vec{\rm U} points towards the galactic anti-center and \vec{\rm W} aligns with the disk rotation axis. The direction of a sightline is specified by the galactic longitude and latitude $(l, b)$ computed using:
    \begin{equation}
        b=\frac{\pi}{2} - \arccos (\vec{w}\cdot \vec{r})
    \end{equation}
and
    \begin{equation}
        l=\left\{ 
        \begin{array}{l l}
            \pi + \arccos (\vec{u}\cdot \vec{r}_{uv}) &\  \text{if}\ 0\leq\arccos(\vec{v}\cdot\vec{r})<\frac{\pi}{2}  \\
            \pi - \arccos (\vec{u}\cdot \vec{r}_{uv}) &\  \text{if}\ \frac{\pi}{2}\leq\arccos(\vec{v}\cdot\vec{r})\leq\pi, 
        \end{array} \right.
    \end{equation}
where \vec{r}$_{uv}$ is the sightline vector \vec{r} projected onto the disk \vec{\rm UV} plane, and \vec{u}, \vec{v},  \vec{w} are unit vectors along the \vec{\rm UVW} axes, respectively.

\begin{figure}[t!]
    \centering
        \includegraphics[trim=0mm 0mm 0mm 1.5cm, clip,width=0.5\textwidth]{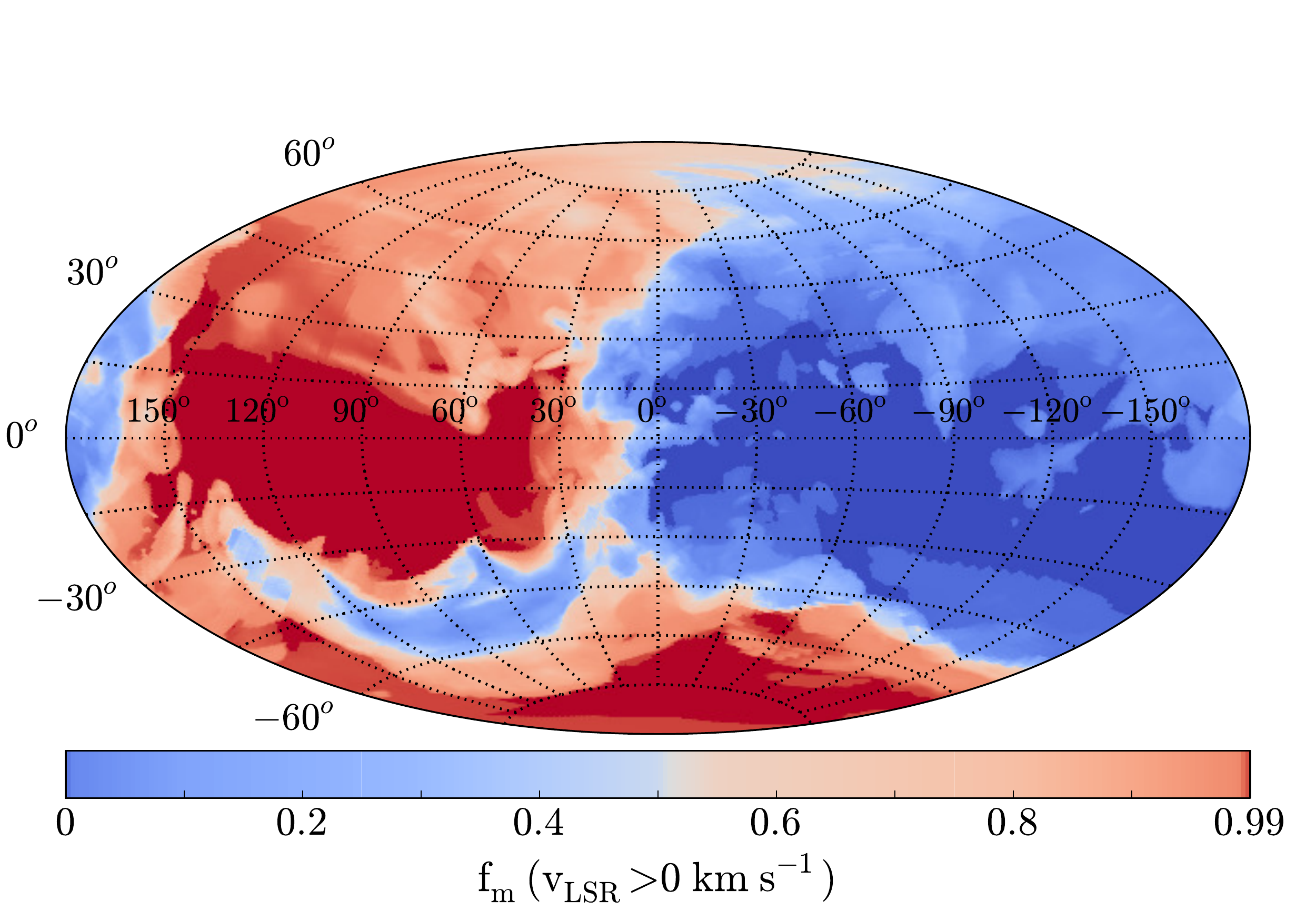}
        \caption{Mass fraction of red-shifted gas ($\vlsr>0$) in the simulated CGM. The map is in Aitoff projection with galactic center at the center of the figure and galactic longitude increasing leftwards. Due to the circular motion of the LSR, the sky is heavily red-shifted in directions at  0$^{\degree}<l<$180$^{\degree}$ and blue-shifted at -180$^{\degree}<l<$0$^{\degree}$. The diminishing color scales towards higher latitudes indicate the existance of both red-shifted and blue-shifted gas.}
    \label{fig:f1} 
    \end{figure}

The mock observer co-rotates with the simulated disk, and its circular velocity is approximated by the bulk motion of baryons within 1.0 kpc of the LSR. To visualize the overall gas motions in the simulated CGM with respect to the mock observer, we generate sightlines with a spatial interval of $(\delta l, \delta b)=(1^{\degree}, 1^{\degree})$ and calculate the mass fraction of red-shifted gas ($\vlsr>$0) along each sightline. Figure \ref{fig:f1} shows the spatial variation of the mass ratio in Aitoff projection, with the galactic center located at the center of the figure and galactic longitude increasing leftwards. Velocity segregation across $l$=180$^{\degree}$ can be observed: gas at 0$^{\degree}<l<$180$^{\degree}$ is largely red-shifted and that at -180$^{\degree}<l<$0$^{\degree}$ blue-shifted. This is consistent with observations of the MW's CGM whose velocities are greatly shaped by the Galactic differential rotation in the view of local observers (e.g, \citealt{Wakker91c}, \citealt{Sembach03}, \citealt{Collins05}, \citealt{Shull09}).

    \begin{figure}[t!]
        \centering
        \includegraphics[trim=8mm 0.5cm 0mm 0cm, clip, width=0.5\textwidth]{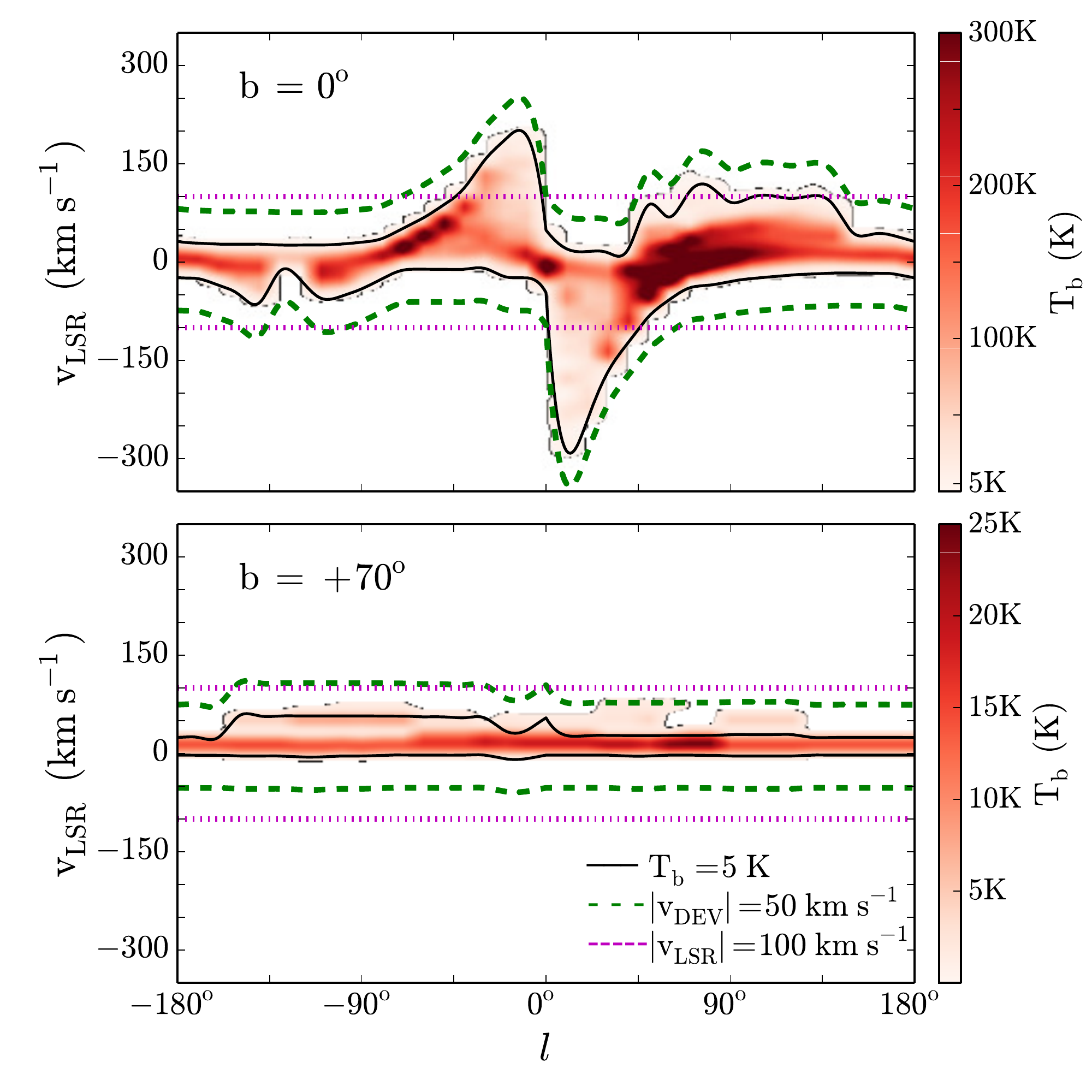}
        \caption{\ion{H}{1} intensity as a function of LSR velocity and galactic longitude ($l$) at $b=0^o$ (top) and $b=70^o$ (bottom). The black solid curves represent the T$_b=5$ K contour used to set the galactic rotation limits $(v_{min}, v_{max})$ in this paper.  The green dashed curves show the $|\vdev|=50\kms$ dividing lines beyond which the gas in the simulated CGM is observable and the horizontal magenta lines show the $|\vlsr|=100\kms$ dividing lines.} 
    \label{fig:f2} 
    \end{figure}

\subsection{Kinematic Ranges of the Simulated Disk Rotation}
\label{sec3.2}
The strong emission or absorption from the disk gas has long plagued observations of the MW's CGM. In most directions the Galactic disk gas appears as low-velocity spectral line features that can be efficiently excluded with certain velocity cutoffs. The adopted velocity range for the gas associated with the MW disk is often $|\vlsr|<90-100\kms$ \citep{Wakker91b, Wakker97,Sembach03,Lehner12} where $\vlsr$ is the velocity with respect to the LSR. A more accurate value is $|\vdev|<50-75\kms$ (\citealt{Wakker91b}, PPJ12) where $\vdev$ is the deviation velocity given by \cite{Wakker91b} as,

    \begin{equation}
        \vdev = \left\{ 
        \begin{array}{l l}
            \vlsr - \vmin & \quad \text{if}\ \vlsr<0  \\
            \vlsr - \vmax & \quad \text{if}\ \vlsr\geq0
        \end{array} \right .
    \end{equation}
. The $(v_{min}, v_{max})$ above is the minimum and maximum LSR velocities compatible with the Galactic differential rotation in a given direction.  This can be obtained either by assuming a simple Galactic rotation model or tracking the motions of neutral hydrogen or molecular clouds in the disk \citep{Hartmann97, Dame01}.

In our synthetic observations, the $(v_{min}, v_{max})$ is calculated from the synthesized \ion{H}{1} 21 cm emission line from the simulated disk (following the description in \citealt{Draine11}; also see \citealt{Kim14}). We approximate the $(v_{min},\ v_{max})$ by the velocities beyond which the \ion{H}{1} brightness temperature T$_b$ drops below 5 K in an \ion{H}{1} spectrum. Figure \ref{fig:f2} shows two \ion{H}{1} position-velocity maps at $b=0^{\degree}$ and $b=70^{\degree}$, respectively. The 5 K temperature threshold defining the $(v_{min},\ v_{max})$ is marked by two black solid curves. As seen in the top panel, the $(v_{min}, v_{max})$ are very direction-sensitive which indicates the strong impact of the circular motion of the LSR on the observed radial velocities at low galactic latitudes. In the simulated disk, a small warp is found in the $\vec{\rm +V}$ direction bending towards the $\vec{\rm -W}$ axis and slightly goes beyond the surface of the disk cylindrical region (also see Figure \ref{fig:f7}). This causes the asymmetry of the \ion{H}{1} 21cm intensities at $l=\pm90^o$ in top panel. \ion{H}{1} intensities below 5 K do not extend significantly, indicating that this cutoff encompasses most of the \ion{H}{1} 21 cm emission from the simulated disk. 

Using $(v_{min},\ v_{max})$, we define two velocity cutoffs to mask out the low velocity gas in the simulated disk: $|\vdev|\geq50\kms$, which closely traces the shape of the galactic rotation in each direction; and $|\vlsr|\geq100\kms$, which is insensitive to direction but also independent of disk rotation models. Gas with velocities beyond (within) these ranges will be referred as ``high (low) velocity gas" throughout the paper. In Figure \ref{fig:f2}, the $|\vlsr|=100\kms$ and $|\vdev|=50\kms$ velocity cutoffs are shown as the dotted magenta lines and the dashed green curves. The velocity cutoff of $|\vlsr|\geq100\kms$ fails to cover high velocity gas near the galactic center which is still associated with the disk, while in other directions it is so wide that excess gas in the simulated CGM is excluded.

\section{RESULTS}
\label{sec4}

\subsection{Kinematic Similarities: \\ Simulated CGM vs. L$\sim$L$_*$ Galaxies}
\label{sec4.1}
    
    \begin{figure}[t!]
     \centering
         \includegraphics[width=0.5\textwidth]{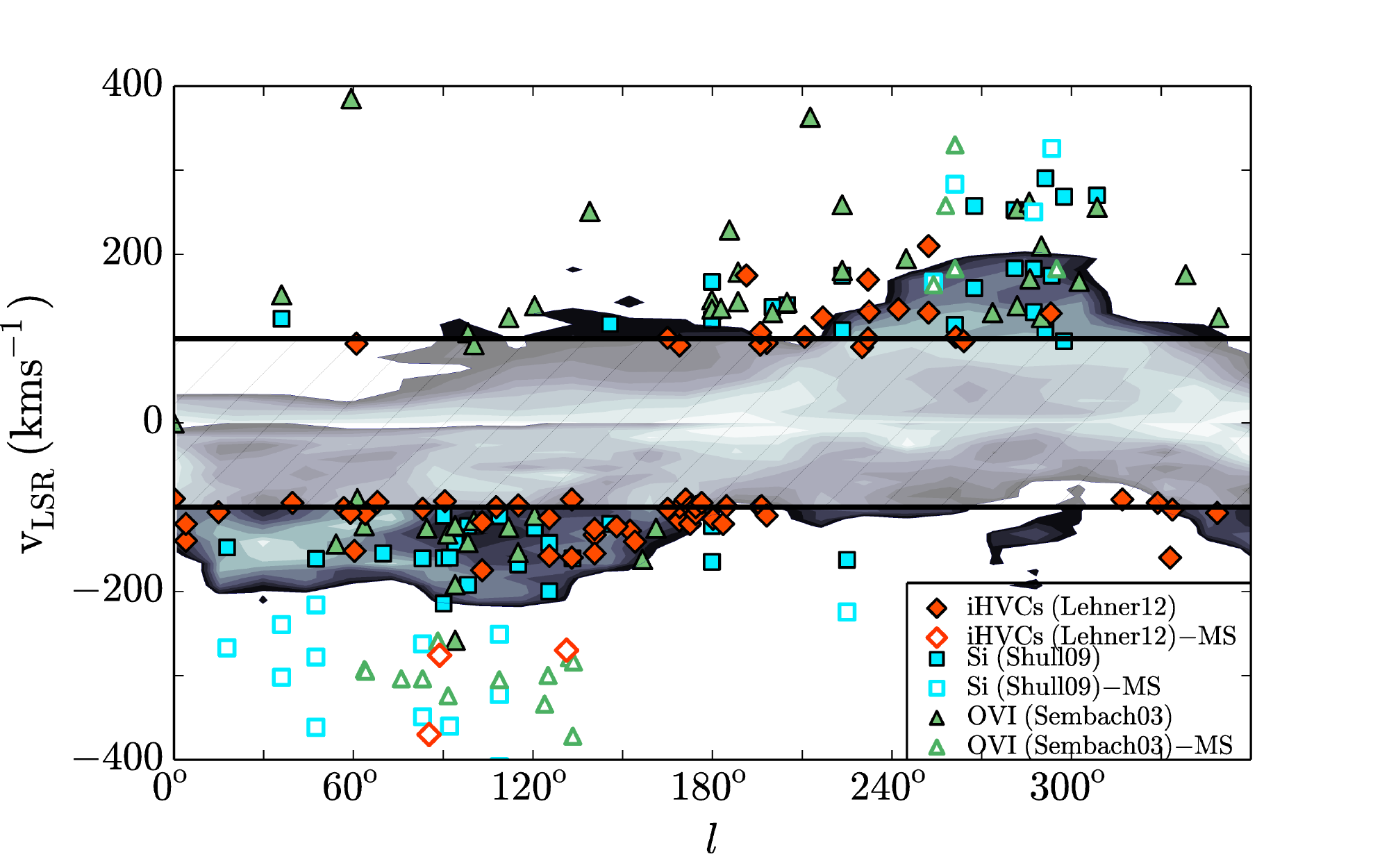}
         \caption{The position-velocity(LSR) distribution of the CGM (T$\leq$10$^6$ K) when observed from within the simulated disk is shown by the shaded region. The mass-weighted LSR velocities of the simulated CGM gas are shown, with lighter shades indicating a higher density of material. The velocity cutoff at $|\vlsr|=100\kms$ is shown by the horizontal black lines and the simulated CGM within this cutoff is lightened. Color-coded symbols show the velocity centroids of high-velocity gas in the MW's CGM from observations (\citealt{Lehner12}, \citealt{Shull09} and \citealt{Sembach03}). Open symbols indicate the association of the observed high-velocity gas with the Magellanic System (MS). \\ - Plot generated with the software {\it astroML} \citep{astroML}.}
     \label{fig:f3}
     \end{figure}
    
We first compare the kinematic properties of the simulated CGM (T$\leq$10$^6$ K) and that observed for the MW. For each sightline through the simulated CGM, we calculate the mass-weighted LSR velocity ($v_{\rm LSR}$) over all of the simulated gas cells as shown in Figure \ref{fig:f3}. Observations of high velocity gas in the MW's CGM from various metal lines are shown as color-coded symbols (e.g., \ion{Si}{2}, \ion{Si}{3}, and \ion{O}{6}; \citealt{Sembach03, Shull09, Lehner12}), and shades represent the simulated CGM. When the observed CGM gas associated with the Magellanic System is excluded (open symbols at generally $|\vlsr|\geq$200$\kms$), there is overall consistency in the gas kinematics between the simulated CGM and the MW's CGM. The velocity spread of the gas in the simulated CGM is narrower than that of the actual MW. This is most likely due to the mass-weighted algorithm which smooths out high-velocity low-density features. Though the simulated MW does have small satellites, they are not at the scale of the Magellanic System and contribute negligible mass flux \citep{Joung12}. In addition, three of four satellites in the simulation at z$\sim$0 are at velocities $\leq$200$\kms$ with respect to the galaxy systemic velocity.

    \begin{figure}[t!]
    \centering
        \includegraphics[width=0.5\textwidth]{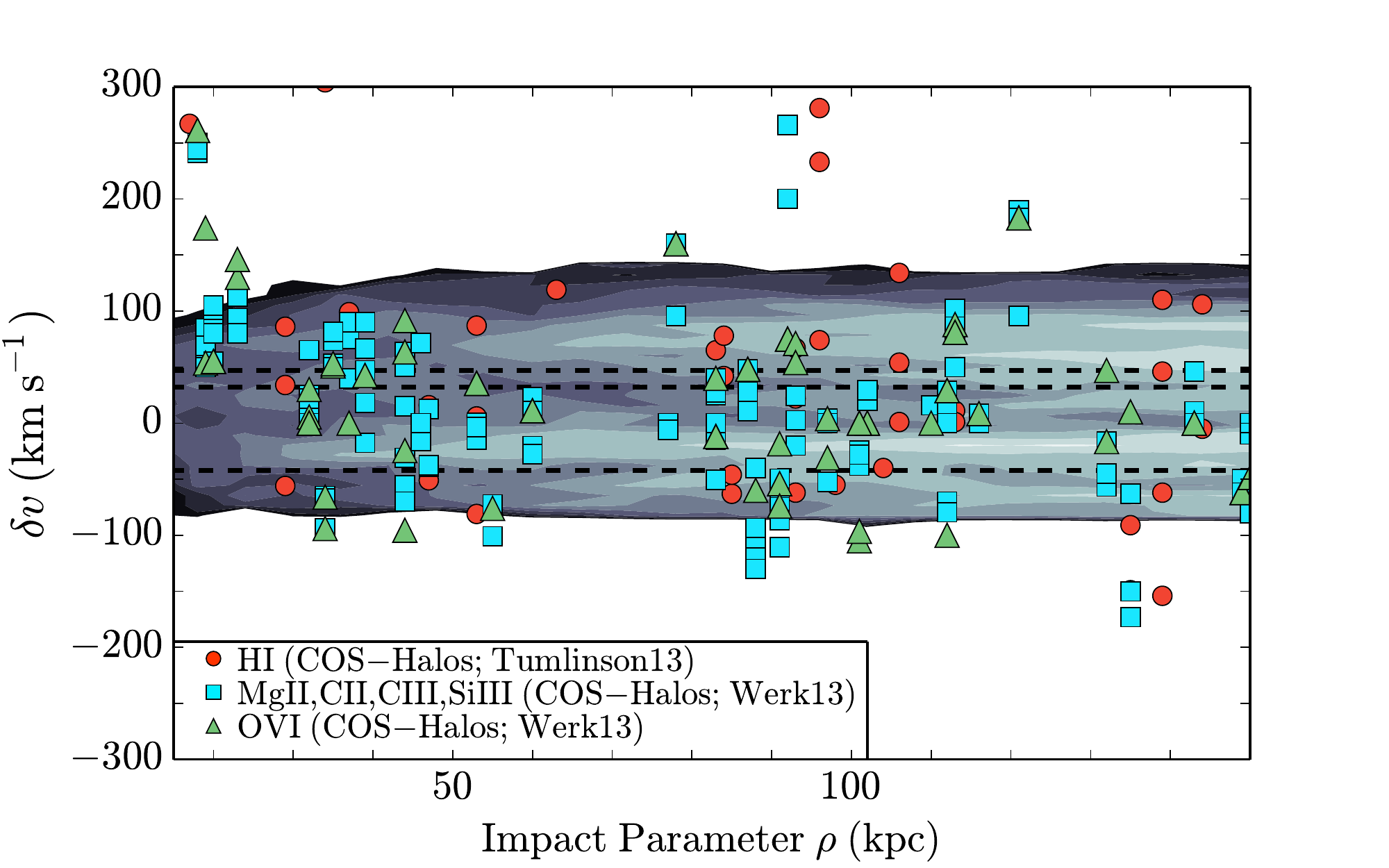}
        \caption{The velocities of the CGM (T$\leq$10$^6$ K) with respect to the galaxy systemic velocity as a function of impact parameter. The gray shades show the line-of-sight mass-weighted velocities of the gas in the simulated CGM when observed from three external views: edge-on, 45$^{\degree}$ and face-on. The three horizontal dash lines indicate the median values of the gas velocities from the edge-on view (bottom), 45$^{\degree}$ (middle) and face-on view (top). Color-coded symbols show the gas centroid velocities of different transition lines from the COS-Halos survey \citep{Tumlinson13, Werk13}. Multiple data points at the same impact parameters represent the multiple component structures along the corresponding line of sight which results from individual Voigt profile component analysis. \\ - Plot generated with the software {\it astroML} \citep{astroML}.}
    \label{fig:f4}
    \end{figure} 

Synthetic observations of the simulated CGM from external views are also conducted so that we can make connections with other L$\sim$L$_*$ galaxies. We observed the simulated CGM (T$\leq$10$^6$ K) with three disk inclinations: $0^{\degree}$ (face-on), $45^{\degree}$, and $90^{\degree}$ (edge-on). From each inclination, we calculate the line of sight velocities for $500\times500$ sightlines on a square grid that spans -250 kpc to 250 kpc (with intervals of 1 kpc) from the galactic center. In Figure \ref{fig:f4}, the line of sight velocity $\delta v$ with respect to the galaxy systemic velocity as a function of impact parameter is plotted as the gray shades for the simulated CGM. The observation symbols are from the COS-Halos survey \citep{Tumlinson13, Werk13} and are similarly color and shape-coded as Figure \ref{fig:f3} showing the observed velocity centroids of different species (i.e., \ion{H}{1}, \ion{Mg}{2}, \ion{C}{2}, \ion{C}{3}, \ion{Si}{3} and \ion{O}{6}) in the CGM of L$\sim$L$_*$ galaxies at z$\sim$0.2. The velocity dispersion of the simulated CGM is $\sigma\approx53\kms$, which is smaller than that derived from the COS-Halos survey ($\sigma\approx85\kms$ from the metal lines by \citealt{Werk13} and $\sigma\approx200\kms$ in the \ion{H}{1} by \citealt{Tumlinson13}). Some of the difference could be the lack of AGN feedback in the simulation, but more important effect is likely that only one simulated galaxy is examined in our synthetic observations while in the COS-Halos' sample galaxies may slightly vary in mass, gas content and so on. High velocity outliers can be seen from the \ion{H}{1} data \citep{Tumlinson13}, that may come from potential fast moving satellites in the CGM of the L$\sim$L$_*$ galaxies or non-relevant cold gas parcels beyond the CGM but intersected by the sightlines. Overall, the kinematics of gas in the simulated CGM when viewed externally is in good agreement with the observations of the CGM of other L$\sim$L$_*$ galaxies.

    \begin{figure}[t!]
    \centering
        \includegraphics[trim=0mm 2cm 0mm 4cm, clip, width=0.5\textwidth]{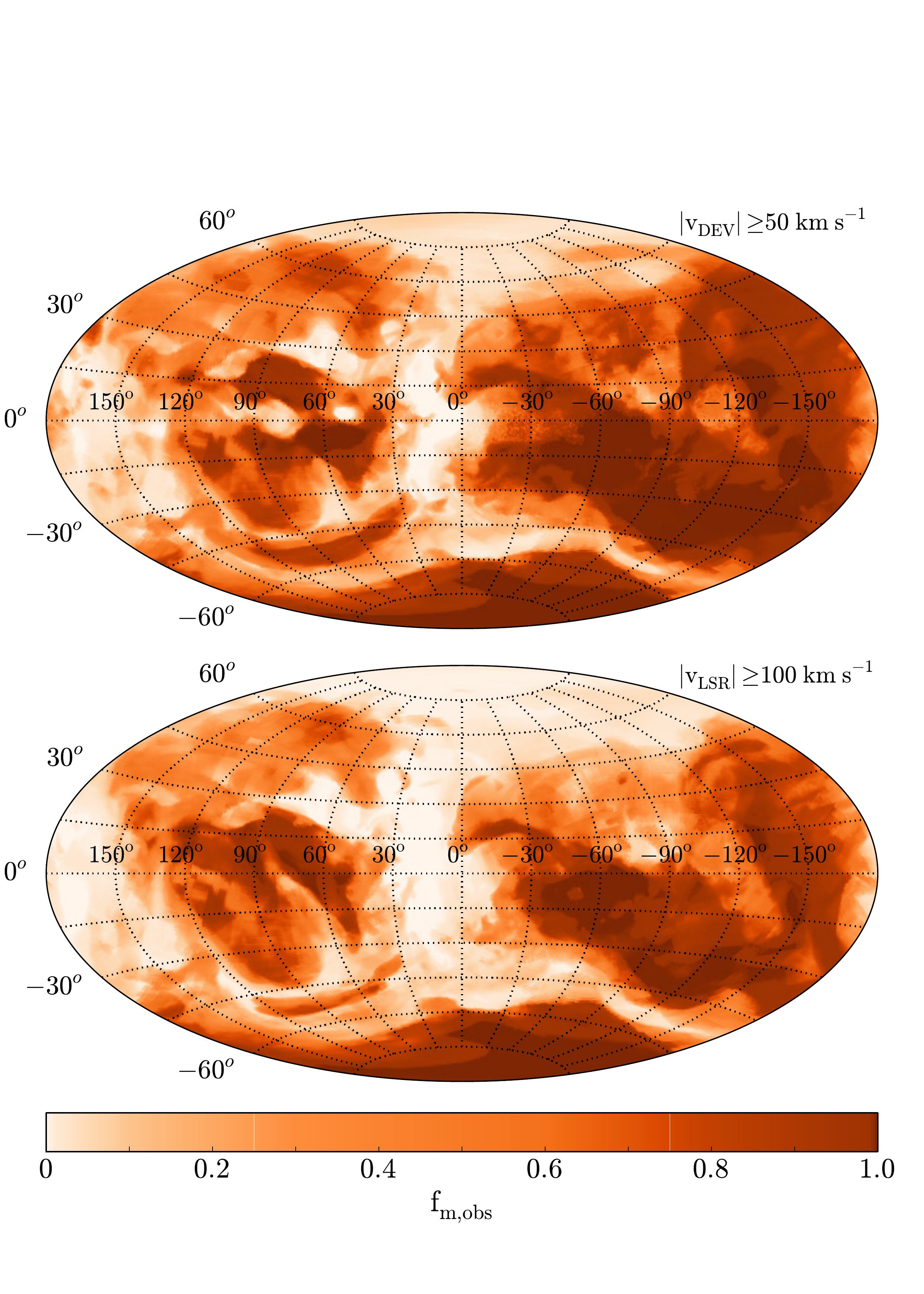}
        \caption{Observable mass fraction $f_m$ of the simulated CGM with velocity cutoffs $|\vdev|\geq50\kms$ (top) and $|\vlsr|\geq100\kms$ (bottom). Figures are in the same configuration as Figure \ref{fig:f1}. Sightlines that contain primarily high-velocity CGM are the darkest, while regions where the CGM is largely obscured are the lightest.}
    \label{fig:f5} 
    \end{figure} 

When viewed from different inclinations, the mass-weighted velocity of the CGM gas shows asymmetric distributions. This is indicated by the dashed lines in Figure \ref{fig:f4}, which are the median values of the velocity distributions for edge-on (bottom), 45 degree (middle) and face-on (top) views. We find a bulk motion of the simulated CGM with respect to the entire galaxy in a direction parallel to the disk plane with $\delta v$ of 30$\kms$. This bulk flow therefore contributes to the line of sight velocities of the simulated CGM from the edge-on view, while observations with the face-on view will largely miss it. This is consistent with \cite{Kacprzak10} who showed that the \ion{Mg}{2} absorption line profiles in their sample favor one side of the galaxy systemic velocity, suggesting that the observed velocity spreads may depend on galaxy inclinations. 
    
\subsection{Gas Observability in the Simulated CGM}
\label{sec4.2}

\subsubsection{Overall Gas Observability}
\label{sec4.2.1}
\begin{figure*}[t!]
    \centering
        \includegraphics[trim = 0mm 8mm 0mm 1cm, clip,width=\textwidth]{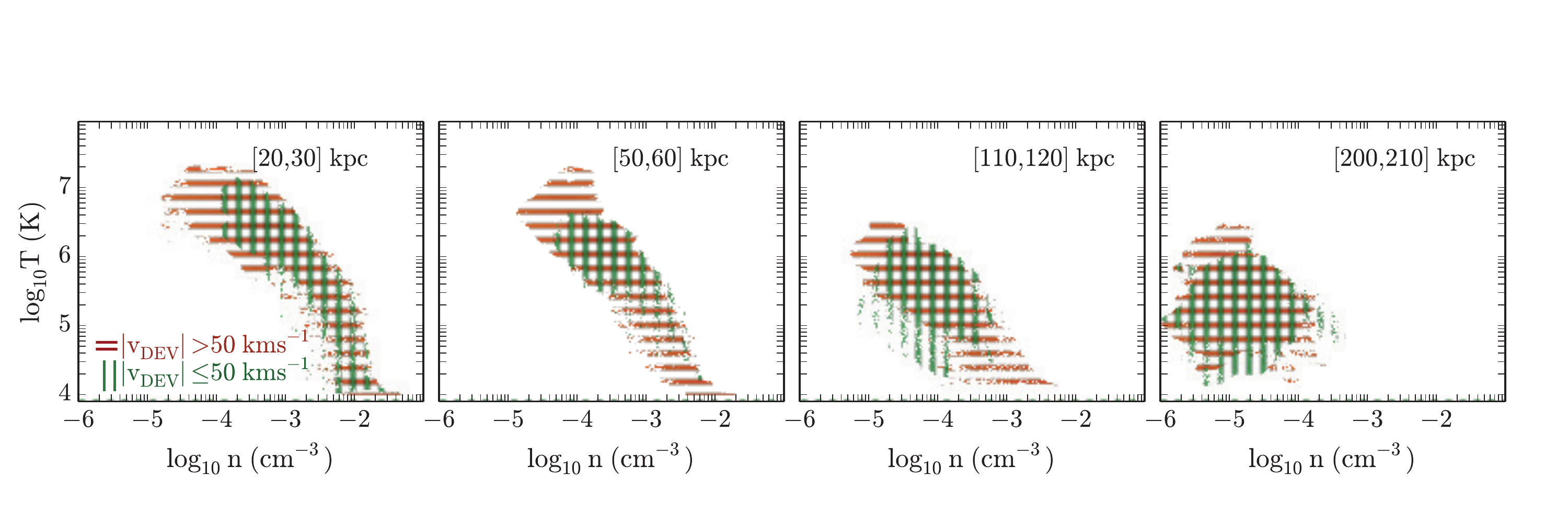}
        \caption{Phase diagrams of the simulated CGM in galactocentric shells at $[20,30]$, $[50,60]$, $[110, 120]$, $[200, 210]$ kpc. High velocity gas moving at $|\vdev|\geq50\kms$ (i.e., observable) is shown by horizontal red strips, while that with $|\vdev|<50\kms$ is marked by vertical green.}
    \label{fig:f6} 
    \end{figure*}

We first calculate the mass fraction and spatial distribution of high velocity gas in the simulated CGM. For a given sightline ($l$, $b$), the mass fraction $f_m$ of high velocity gas along a path length of 250 kpc is computed. The top (bottom) panel in Figure \ref{fig:f5} shows the variation of $f_m$ across the simulated galactic sky with a velocity cutoff of $|\vdev|\geq50\kms$ ($|\vlsr|\geq100\kms$). Close to $l\pm$90$^{\degree}$, large patches with $f_m\rightarrow1$ are observed. This indicates that most gas along corresponding sightlines move at high velocities which are observable to the mock observer. As mentioned in Section \ref{sec3.2}, the velocity cutoff $|\vlsr|\geq100\kms$ is generally too broad for most directions, resulting in more obscuration of the simulated CGM. This reflects in Figure \ref{fig:f5} (and Table \ref{tb:fm}), where $f_m$ with $|\vlsr|\geq100\kms$ is generally lower than that with $|\vdev|\geq50\kms$. At the bottom of both panels exists a large orange patch, resulting from high velocity SN-driven outflows from the simulated disk \citep{Joung12}. This outflowing gas can also be found in the phase diagrams (Figure \ref{fig:f6}) at temperature above 10$^6$ K. 

    \begin{deluxetable}{cccc}
    \tabletypesize{\footnotesize} \tablewidth{0pt}
    \tablecaption{Observability}
    \tablehead{
    \colhead{Components} &
    \colhead{Norm. Mass} & 
    \colhead{$f_{m, \vdev}$} & 
    \colhead{$f_{m, \vlsr}$} \\
    \colhead{(1)} &
    \colhead{(2)} &
    \colhead{(3)} & 
    \colhead{(4)} 
    }
    \startdata
    CGM (r$\leq$250 kpc) & 1.00 & 0.54 & 0.45 \\
    \hline
    $|b|\geq$10$^{\degree}$ & -- & 0.44 & 0.40 \\
    $|b|\geq$20$^{\degree}$ & -- & 0.33 & 0.35 \\
    \hline
    $T<10^5$K & 0.08 & 0.62 & 0.55 \\
    $10^5\leq T<10^6$K & 0.70 & 0.54 & 0.44  \\ 
    $T\geq10^6$K & 0.22 & 0.53 & 0.44   
    \enddata
    \tablecomments{Column 2: gas mass normalized to the total mass of the halo; column 3: mass fraction of high velocity gas with $|\vdev|\geq50\kms$; column 4: gas mass fraction with $|\vlsr|\geq100\kms$. Row 2: overall gas observability in the simulated CGM; row 3: observability of halo gas at $|b|\geq10^{\degree}$ and $|b|\geq20^{\degree}$; row 4: gas observability within temperature ranges $T<10^5$ K, $10^5\leq T<10^6$ K and $T\geq 10^6$ K, respectively.}
    \label{tb:fm}
    \end{deluxetable}

Integrating $f_m$ over the whole spherical volume ($r\leq$250 kpc), we obtain the total mass fraction of high velocity gas for different CGM components defined by temperature ranges and galactic latitudes. In Table \ref{tb:fm}, column 2 shows the mass of each component normalized to the total gas mass of the simulated CGM. The $f_{m, \vdev}$ ($f_{m, \vlsr}$) in column 3 (4) indicates the mass fraction of high velocity gas with the velocity cutoff $|\vdev|\geq50\kms$ ($|\vlsr|\geq100\kms$). \emph{We find that generally only $\sim$0.5 of the CGM mass is carried by high velocity gas; this means that nearly half of the mass of the simulated CGM is obscured from observations due to velocity cutoffs}. This is one of the key results of this paper. 

Specifically, the mass fractions at $|b|\geq10^{\degree}$ and $|b|\geq20^{\degree}$ are computed since low latitude regions in the actual MW's CGM are usually avoided by absorption line experiments due to Galactic emission or absorption. We find that the $f_{m, \vdev}$ ($f_{m, \vlsr}$) decreases to $\sim$0.4 at $|b|\geq10^{\degree}$ ($\sim$0.3 at $|b|\geq20^{\degree}$), indicating that the observations of the actual MW's CGM may miss more than half of the gas mass. It is notable from Figure \ref{fig:f5} that the simulated southern sky is covered by more high velocity gas than the northern sky. The MW (and other actual galaxies) may have this type of imbalance but it is biased by the presence of the Magellanic System in the southern sky. 

\subsubsection{Gas Observability at Various Phases and Galactocentric Radii}
\label{sec4.2.2}
In order to study if velocity cutoffs preferentially select gas at certain phases, we calculate the mass fraction of high velocity gas for gas with temperature $T<10^5$ K, $10^5{\rm K}\leq T<10^6$ K and $T\geq10^6$ K. The last row in Table \ref{tb:fm} shows that the mass fraction of high velocity gas remains nearly 50$\%$ for gas at different phases in spite of their distinct contributions to the total CGM mass. We compare the phase properties of the high velocity and low velocity gas in the simulated CGM in Figure \ref{fig:f6}, and find that they occupy similar regions in the phase space (see below). This indicates that observations of the MW's CGM using various metal lines from different ionization states may miss similar amounts of low velocity gas no matter which temperature ranges they probe. 

The \ion{H}{1} HVCs in the actual MW's CGM are generally observed within 0.06 R$_{\rm vir, MW}$ while the COS-Halos team finds CGM out to 0.55 R$_{\rm vir, COS}$ \footnote{We adopt R$_{\rm vir, MW}$ as 250 kpc, same as the value we used for our synthetic observations; and R$_{\rm vir, COS}\approx$300 kpc is given by the COS-halo teams \citep{Werk14}.}. To study if velocity cutoffs preferentially select high velocity gas at certain radii, we decompose the volume of the simulated CGM into a set of galactocentric shells, each with a thickness of $\Delta r$=10 kpc. We separately calculate the mass fraction of high velocity gas within each shell, and find that $f_m$ fluctuates very little from $r$=10 kpc to $r$=250 kpc, with an average of 0.54$\pm$0.03 for $|\vdev|\geq50\kms$ (0.44$\pm$0.03 for $|\vlsr|\geq100\kms$). Such constant $f_m$ throughout the simulated CGM can be explained if the gas velocities of the simulated CGM are relatively continuous along radial directions (see Section \ref{sec4.4} for further discussion). 

Figure \ref{fig:f6} shows the phase of high and low velocity gas within the different galactocentric shells. Only the result with $|\vdev|\geq50\kms$ is shown since that with $|\vlsr|\geq100\kms$ is similar. Though the high and low velocity gas in the simulated CGM do generally occupy similar phase space, the high velocity gas is more extended to the high-temperature, low-density region (the top left corner) which corresponds to high velocity, hot outflows in the simulation (\citealt{Joung12}, PPJ12). There is also some high density, low temperature gas in the shells between 50-120 kpc that is only found in the high velocity regime. This corresponds to gas stripped from recently accreted satellites \citep{Joung12} and is a negligible fraction of the total CGM mass. We also examined the simulated CGM at different metallicity bins ranging from 0-0.1 Z$\odot$ to $>$1 Z$\odot$, and find that the mass fractions of high velocity gas are similar ($f_m$ $\sim$0.6) in spite of the distinct mass distribution at each bin. These results suggest that we are not biased towards high velocity gas (${\rm T\leq10^6\ K}$) having specific properties that may represent distinct origins. 

Within each galactocentric shell, high velocity gas can generally be found at directions towards $l\pm$90$^{\degree}$ which is similar to the spatial distribution of the overall case shown in Figure \ref{fig:f5}. This is to say, the velocity fields in the simulated CGM are closely correlated in space throughout the volume. This spatial coherence is in large part due to the circular motion of the LSR, which shifts halo gas around $l\pm$90$^{\degree}$ to high velocity by adding extra components to the gas radial velocity with respect to the LSR. On the other hand, such large scale velocity correlation is also aided by the presence of large scale gas flows, such as feeding filaments and polar outflows. 

\subsubsection{Varying the Location of the Mock Observer}
\label{sec4.2.3}
Gas observability in the simulated CGM discussed above can be biased since the mock observer is so far fixed at one location in the disk mid-plane. Local activities such as turbulence and SN-driven outflows may cause non-negligible effects on the corresponding synthetic observations. In this section, we vary the location of the mock observer in the disk mid-plane and study if the change of the reference LSR affects the observed gas kinematics in the simulated CGM. 

We relocate the mock observer to seven other places in the simulated disk with an angular separation of $\phi=45^{\degree}$, where $\phi$ is the polar angle on the fiducial $\vec{\rm UV}$ plane defined in Section \ref{sec3} with $+\vec{\rm U}$ being the polar axis. Each location is 8.0 kpc from the galactic center and the velocity of a given LSR is again averaged over the baryons in its local neighbourhood in order to closely represent the circular motion of the disk rotation at that location. We follow the same procedures described in Section \ref{sec4.2.1} and calculate the mass fraction of high velocity gas in the simulated CGM from corresponding new locations. We only apply $|\vlsr|\geq100\kms$ in our calculations but note that the values with $|\vdev|\geq50\kms$ will be similar considering the results in Section \ref{sec4.2.1}. 

Table \ref{tb:lsr} shows the mass fraction $f_{m, v_{\rm LSR}}$ of high velocity gas in the simulated CGM as observed from new locations. $f_{m, v_{\rm LSR}}$=0.45 at $\phi$=0$^{\degree}$ corresponds to the value from the original LSR, which has been shown in Table \ref{tb:fm}. The gas observability in the simulated CGM varies from 0.39 to 0.75 (with a mean of 0.55) as the angular location of the mock observer changes. As indicated in Section \ref{sec4.1}, a net bulk flow ($\delta v\sim$30$\kms$) of the simulated CGM with respect to the galaxy is observed in our simulation. This may largely cause the variation of $f_{m, v_{\rm LSR}}$ since the calculations of the velocities of the LSR only consider the bulk motions of local neighbourhood. 

We examine the spatial distribution of high velocity gas in the simulated CGM as observed from these eight different locations, and find similar high velocity patterns as have been seen in Figure \ref{fig:f5}. Except for the hot outflowing gas towards the poles, most high velocity gas is observed in directions around $l\pm$90$^{\degree}$, indicating that gas velocities at low galactic latitudes are greatly shaped by the circular motion of the LSR in the disk (see also Figure \ref{fig:f1}). In all cases, we still find that nearly half of the CGM mass is carried by low velocity gas, which is obscured due to velocity cutoffs in spite of different locations of mock observers in the disk plane. 

    \begin{deluxetable}{ccccccccc}
    \tabletypesize{\footnotesize} \tablewidth{0pt}
    \tablecaption{Variation of Observability}
    \tablehead{
    \colhead{$\phi$} &
    \colhead{0$^{\degree}$} & 
    \colhead{45$^{\degree}$} & 
    \colhead{90$^{\degree}$} &
    \colhead{135$^{\degree}$} &
    \colhead{180$^{\degree}$} &
    \colhead{225$^{\degree}$} & 
    \colhead{270$^{\degree}$} &
    \colhead{315$^{\degree}$} 
    }
    \startdata
    $f_{m, v_{\rm LSR}}$ & 0.45 & 0.39 & 0.55 & 0.70 & 0.67 & 0.64 & 0.53 & 0.45
    \enddata
    \tablecomments{Mass fraction of high velocity gas as a function of the locations of the mock observer. Velocity cutoff is set as $|\vlsr|\geq100\kms$. Each location is 8.0 kpc from the galactic center, and $\phi=45^{\degree}$ apart where $\phi$ is the polar angle on the $\vec{\rm UV}$ plane defined in Section \ref{sec3}.}
    \label{tb:lsr}
    \end{deluxetable}

\subsection{{\rm \ion{H}{1}} Observability in the Simulated CGM}
\label{sec4.3}

In this section, we calculate the distribution and covering fraction of high velocity \ion{H}{1} in the simulated CGM. We only present the \ion{H}{1} behaviors with velocity cutoff $|\vdev|\geq50\kms$, with the knowledge that the results with $|\vlsr|\geq100\kms$ are similar according to Figure \ref{fig:f5}. For a given sightline, we compute the column density of high velocity \ion{H}{1} ($\Nhi$) by summing up the \ion{H}{1} retained by all the simulated gas cells with $|\vdev|\geq50\kms$ along the corresponding path ($\sim$250 kpc). We also use different path lengths ranging from 50 to 300 kpc and find little variations among those $\Nhi$ values we have obtained. This is because most of the \ion{H}{1} resides at radii close to the disk except for satellites, an increase in path length will not necessarily gather more \ion{H}{1}.

\begin{figure}[t!]
    \centering
        \includegraphics[trim=0mm 0mm 0mm 1.5cm, clip,width=0.5\textwidth]{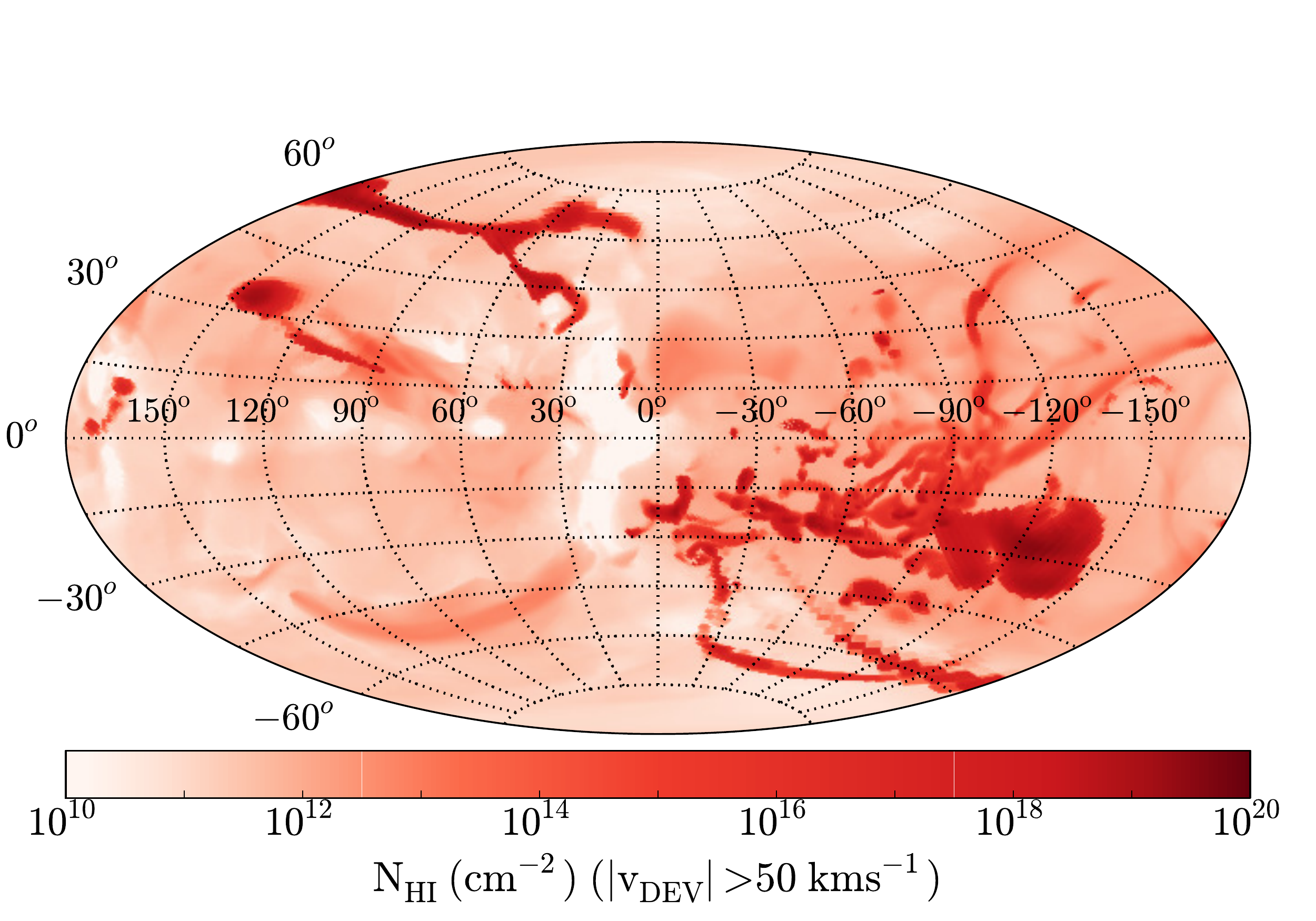}
        \caption{Column density of high velocity \ion{H}{1} ($|\vdev|>50\kms$) in Aitoff projection. The galactic center is located at the center of the figure, and galactic longitude $l$ increases from left to right.}
    \label{fig:f7}
    \end{figure}

Figure \ref{fig:f7} shows the spatial distribution of high velocity \ion{H}{1} in the simulated CGM. The high $\Nhi$ clumps near $(l,b)=(-90^{\degree}, -30^{\degree})$ are the warp of the disk bending downwards which was not included in the cylindrical disk region as defined in Section \ref{sec3.2} (see also \citealt{Fernandez12}). This uneven distribution of \ion{H}{1} at $l=\pm90^{\degree}$ causes the asymmetric \ion{H}{1} 21 cm emission in Figure \ref{fig:f2}. At higher elevation above the disk plane, elongated structures can be seen which may relate to gas-rich satellites or accreted gas along filamentary streams in the simulated CGM \citep{Joung12, Fernandez12}. Besides these distinct dense structures, most of the simulated galactic sky is covered by very diffuse \ion{H}{1}. 

\begin{figure}[t!]
        \centering
        \includegraphics[trim=1cm 0mm 1.5cm 1.3cm, clip, width=0.5\textwidth]{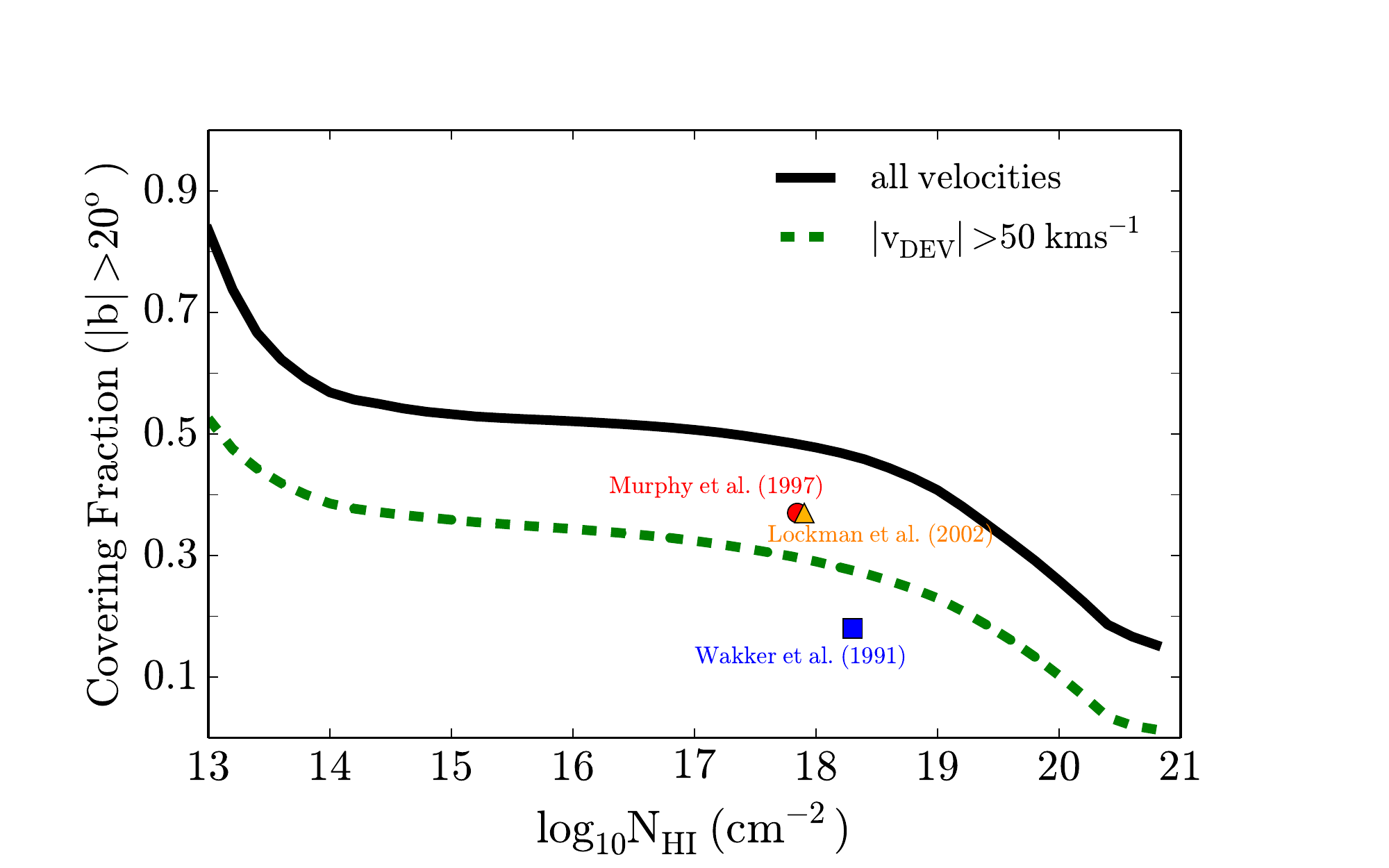}
        \caption{\ion{H}{1} covering fractions vs. \ion{H}{1} column sensitivities in the simulated CGM ($|b|\geq20^{\degree}$). Black solid line: \ion{H}{1} without any velocity cutoffs; green dashed line: \ion{H}{1} with $|\vdev|\geq50\kms$. Color-coded symbols: observed \ion{H}{1} HVCs covering fractions in the actual MW's CGM by \cite{Wakker91b}, \cite{Lockman02} and \cite{Murphy95}.}
    \label{fig:f8} 
    \end{figure}

Figure \ref{fig:f8} shows the covering fractions of high velocity \ion{H}{1} and the \ion{H}{1} without velocity cutoffs in the simulated CGM at different column sensitivities. Regions at $|b|<20^{\degree}$ are purposely omitted in order to link to the actual observations of the MW's CGM. A comparison between the black solid line (without velocity cutoffs) and the green-dash lines (with velocity cutoffs) indicates that the covering fraction of the high velocity \ion{H}{1} is reduced to only 0.29 (from 0.48) at $\Nhi\geq10^{18}$ cm$^{-2}$. More generally, the covering fractions of the high velocity \ion{H}{1} are $\sim$0.2 less than those without velocity cutoffs at a given column density sensitivity. Color-coded symbols at $\Nhi\sim10^{18}$ cm$^{-2}$ mark the measured covering fractions of HVCs in the actual MW's CGM: 0.18 at $\Nhi>2\times10^{18}$ cm$^{-2}$ \citep{Wakker91b}\footnote{No latitude limitation $|b|\geq20^{\degree}$ was adopted by \cite{Wakker91b}.}, 0.37 at $\Nhi>7\times10^{17}$ cm$^{-2}$ \citep{Murphy95}, and 0.37 at $\Nhi>8\times10^{17}$ cm$^{-2}$ \citep{Lockman02}. The consistency between the measured \ion{H}{1} HVCs and the simulated high velocity \ion{H}{1} column density indicates that a considerable fraction of low velocity \ion{H}{1} may remain hidden in the actual MW's CGM (see Section \ref{sec5.3} for a detailed discussion of potential missing low velocity \ion{H}{1} complexes).

\subsection{Additional Mass Obscuration in the Outer Halo}
\label{sec4.4}

Apart from the mass obscuration due to foreground low velocity disk gas, high velocity gas in the outer halo may be additionally blocked by nearby halo gas if it moves at similar velocities. In another words, even at high velocity regimes spectral lines from the gas in the outer halo may blend with signals from nearby gas if line separations (i.e., velocity differences) are too small to detect. Recognition of these close line profiles will depend on the resolution of observations. 

To quantify this issue, we extract two data sets from the simulation which represent an inner halo ($r\leq$30 kpc) and an outer halo (50$\leq r\leq$150 kpc), respectively. Maps of high velocity gas in both halos in Aitoff projection show similar patterns as those in Figure \ref{fig:f5}. For a given sightline, we calculate the velocity differences between the high velocity gas in the inner halo and that in the outer halo. This is separately performed for red-shifted and blue-shifted gas. Gas in the outer halo is observable only if: (1) its velocity is high enough to survive the $|\vdev|\geq50\kms$ or $|\vlsr|\geq100\kms$ cutoffs; (2) its velocity differs from that of the gas in the inner halo by a number $\Delta v$ which is the chosen minimum detectable line separation. 

Table \ref{tb:fm_add} shows the observable mass fraction of high velocity gas in the outer halo that satisfies both criteria under a set of minimum detectable line separations $\Delta v$ ranging from $0\kms$ to $30\kms$. If the spectral resolution is high enough (0$<\Delta v\leq$10$\kms$) so that gas at similar velocities in the inner and outer halo would be distinguishable, we find that the observable mass fraction in the outer halo is 0.47 at $|\vdev|\geq50\kms$ (0.43 at $|\vlsr|\geq100\kms$). However, the observable mass fraction in the outer halo decreases as the spectral resolution falls off (i.e., $\Delta v$ becomes larger). For example, it is significantly reduced by a factor of two (from 0.47 to 0.23) if $\Delta v=30\kms$. This indicates that less than 50$\%$ of gas is accessible to local observers due to a combination effect of velocity cutoffs and obscuration by nearby halo gas. In particular, this has a potential influence on the mass estimate of the ionized HVCs by \cite{Lehner12} who compared the covering fractions from a stellar sample with that from a QSO sample, and constrain the clouds to be within 5-15 kpc. 
     
\section{Discussion}
\label{sec5}

The CGM is key to our understanding of galactic accretion and feedback processes. It has been claimed that the CGM can solve the ``missing baryon problem" from studies with both observations (e.g., \citealt{Werk14}) and simulations (e.g., \citealt{Cen06}, \citealt{Sommer-Larson06}). In this section, we re-evaluate the total mass of the MW's CGM taking into account the obscured mass due to the velocity cutoffs, and compare the obscuration-corrected MW's CGM with the CGM of other L$\sim$L$_*$ galaxies. In the end of this section, we discuss potential observations of obscured low velocity gas in the MW's CGM through non-kinematic methods.

    \begin{deluxetable}{ccccc}
    \tabletypesize{\footnotesize} \tablewidth{0pt}
    \tablecaption{Gas observability in the outer halo \\ 
    ($50\leq r\leq 150\ kpc$)}
    \tablehead{
    \colhead{Min. Sep.} &
    \colhead{$0$} & 
    \colhead{$10$} & 
    \colhead{$20$} & 
    \colhead{$30$} \\
    \colhead{$\Delta v$} &
    \colhead{$\kms$} & 
    \colhead{$\kms$} & 
    \colhead{$\kms$} & 
    \colhead{$\kms$}
    }
    \startdata
    $f_{m, \vdev}$ & 0.47 & 0.39 & 0.30 & 0.23 \\
    $f_{m, \vlsr}$ & 0.43 & 0.31 & 0.23 & 0.16
    \enddata
    \tablecomments{$\Delta v$ means a minimum detectable line separation in spectroscopic observations; $f_{m, \vdev}$ and $f_{m, \vlsr}$ are the observable mass fractions of gas in the outer halo at $|\vdev|\geq50\kms$ ($|\vlsr|\geq100\kms$) with the corresponding $\Delta v$.}
    \label{tb:fm_add}
    \end{deluxetable}

\subsection{Obscuration-corrected Mass of the MW's CGM}
\label{sec5.1}

Cold gas (T$<10^4$ K) in the actual MW's CGM has been broadly studied with \ion{H}{1} 21 cm observations, while the warm ($\sim$10$^{4-5}$ K) and warm-hot gas ($\sim$10$^{5-6}$ K) has been revealed by UV absorption lines in the spectra of background sources (stars or QSOs) and H$\alpha$ emission. Mass estimates of this gas focus on available high velocity lines in spectra, with little evaluation of the mass contribution from low velocity obscured halo gas. Thus in this section we apply the observable mass fraction $f_m$=0.5 to current mass estimates of the MW's CGM and obtain an obscuration-corrected mass for CGM gas at various phases. We do not apply the correction to gas from the Magellanic System as this system is moving at high velocity relative to the MW \citep{Kallivayalil13} and the vast majority of the gas is likely to be observable based on predictions from the models \citep{Besla10, Connors06}. This type of large accretion event is also not presently occurring in our simulated galaxy halo, and is rare in observations of MW-like galaxies \citep{Tollerud11}.

Cold gas in the MW's CGM is usually in the form of discrete \ion{H}{1} structures called HVCs. Galactic HVCs have a total mass of ${\rm M_{HI}}$ $\sim$3$\times$10$^7\Msun$, excluding the HVCs associated with the Magellanic System. This includes 13 \ion{H}{1} complexes within 10 kpc of the disk and the remaining small HVCs placed at a general distance of 10 kpc (PPJ12). Considering the 50$\%$ mass obscuration, the total mass of Galactic HVCs is therefore corrected to ${\rm M_{HI}^{corr}}$ $\sim$6$\times$10$^7\Msun$. This corrected \ion{H}{1} mass in the MW's CGM without the Magellanic System is consistent with observations of M31's halo gas \citep{Thilker04}. If we include the mass from the high velocity gas of the Magellanic System (excluding the Magellanic Clouds themselves), the mass increases by ${\rm M_{HI}=3\times10^8 ({\rm d/55\ kpc})^2 \Msun}$ \citep{Putman03b, Bruns05}.

There are recent mass estimates of warm and warm-hot gas in the MW's CGM. \cite{Lehner11} found that the ionized HVCs reside within 5-15 kpc using the constraint from halo stars. The total mass of warm and warm-hot gas is thus ${\rm M_{warm}}\sim$1.1$\times$10$^8({\rm d/12\ kpc})^2\Msun$ (see also \citealt{Shull09}). If the observable mass fraction $f_m$=0.5 is applied, the mass of warm and warm-hot gas will be ${\rm M_{warm}^{corr}}$ $\sim$2.2$\times$10$^8({\rm d/12\ kpc})^2\Msun$. This mass is almost certainly a lower limit given the consistent finding of warm and warm-hot gas at large radii in the CGM of other galaxies (e.g., \citealt{Tumlinson11, Chen10, Werk14}). Some of the MW's CGM at large radii may be too diffuse to detect from within the disk using background QSOs. One should also consider that in Section \ref{sec4.4} we show that distant gas in the CGM can be obscured by nearby gas if they are moving at similar velocities. This shadowing effects from high velocity gas in the inner halo may cause an additional 50$\%$ of mass obscuration in the outer halo in observations with a spectral resolution of $\geq$30$\kms$. Thus the mass estimate by \cite{Lehner12} shall be interpreted with caution as they do not take into account the local halo shadowing effects. We do not attempt to correct for this additional obscuration here or calculate a mass for gas that does not have a distance constraint. A recent estimate of the total mass in warm and warm-hot gas from the Magellanic System (excluding the Magellanic Clouds) is ${\rm M_{\rm warm, MS}}\approx$1.5$\times$10$^9 ({\rm d/55\ kpc})^2\Msun$ \citep{Fox14}.

\cite{Miller14} estimate the mass of hot gas (T$\gtrsim$10$^6$ K) in the MW's CGM to be 4.3$\times$10$^{10}\Msun$ within 250 kpc (or 3.8$\times$10$^9\Msun$ within 50 kpc) using \ion{O}{7} and \ion{O}{8} lines through X-ray observations (see also \citealt{Gupta12}). This mass is consistent with non-spectroscopic observations (see \citealt{Anderson10} for a detailed discussion). Given the low velocity resolution of X-ray observations, the mass estimates of hot halo gas are usually based on line profile fitting to exclude local emission components rather than applying velocity cutoffs. Thus our mass obscuration correction is not applicable to their mass estimates. In Figure \ref{fig:f6}, we also show that the gas that would be at highest velocities and potentially directly detectable is the most diffuse and hot and the most difficult to detect. 


    \begin{figure*}[t!]
    \centering
        \includegraphics[trim = 0mm 2mm 0mm 4mm, clip,width=\textwidth]{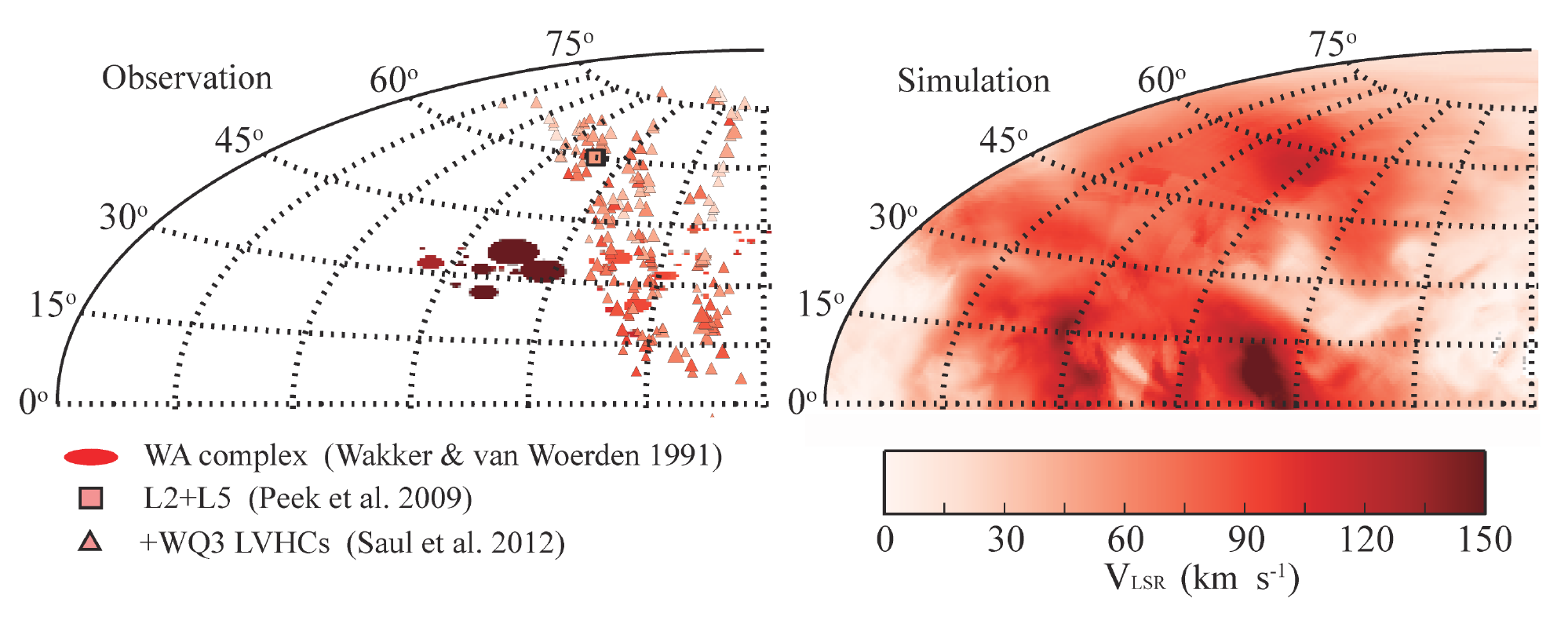}
        \caption{Left: the WA complex from the HVC catalog of \citet{Wakker91c} is shown in ellipses. The size of the ellipse indicates the sky coverage of each WA HVC, but does not correspond to the shape of the observed WA HVCs on the sky. The L2+L5 \citep{Peek09} and +WQ3 LVHCs \citep{Saul12} are indicated by squares and triangles respectively, although their sizes are not accurately indicated. The color of the symbols indicates the LSR velocities. Right: the mass-weighted LSR velocities of the simulated CGM of the same patch of sky. The velocities shown in both panels are in the same color scale as is indicated by the color bar. A velocity gradient can be clearly seen both in the MW and the simulated CGM.}
    \label{fig:f9} 
    \end{figure*}

\subsection{Is the MW's CGM an Outlier?}
\label{sec5.2}
In this section we compare the MW's CGM to that of other L$\sim$L$_*$ galaxies for which large amounts of warm and warm-hot ($10^{4-6}$ K) have been found. We start by looking at the mass estimates and then, given the modeling required to obtain total masses, assess the difference in terms of the column density. As shown in Figure \ref{fig:f4}, external galaxies do not have the same obscuration effect; the observations of background QSOs often show gas in CGM very close to the systemic velocity of the galaxy.

\subsubsection{CGM Mass}
\label{sec5.2.1}
\cite{Stocke13} found $\sim$10$^{10}\Msun$ of CGM (T$\sim$10$^4$ K) within the virial radius of late-type galaxies at z$\leq0.2$ (with 75$\%$ of the mass contained in the inner half of the virial radius). \cite{Werk14} also suggest a conservative estimate of 2$\times$10$^{10}\Msun$ for the 10$^4$ K CGM gas within 160 kpc of the L$\sim$L$_*$ galaxies at z$\sim$0.2. This is $\sim$3$\times$10$^9\Msun$ within 50 kpc of the disk and $\sim$4$\times$10$^8\Msun$ within 15 kpc based on the gas surface density profile they adopt. The value is comparable to the MW estimate (Section \ref{sec5.1}) within small radii, but the MW value is still low and the external galaxy results indicate there is most likely a large amount of gas undetected at large radii.

Since the COS results are for galaxies at somewhat higher redshift, we examined the simulated CGM at different redshifts to assess if any discrepancy could be due to gaseous halo evolution with time. Negligible variation is found for the mass of the simulated CGM from z=0.3 to z=0.0 (within a fixed radius of 250 kpc). This is consistent with \cite{Stinson12} who examined the CGM gas content in an L=1\ L$_*$ star-forming galaxy, and found little variation of CGM mass within the virial radius since z=0.5. Thus, the difference in redshift of the external galaxies and the MW is not expected to greatly affect the CGM mass.


There are some remaining issues with the CGM mass estimates for external galaxies. The warm CGM pressure that \cite{Stocke13} derive (P/k$\sim$10 cm$^{-3}$ K) is an order of magnitude lower than that typically found from observations of clouds in the halo of the Milky Way \citep{Fox10, Hsu11}. This may be due to uncertainties related to the photoionization modeling. In addition, one of the mass estimates for the $10^4$ K CGM by \cite{Werk14} is based on the assumption that the $\nhi$ along each sightline is representative of all of the gas at a given impact parameter. However, observations of the MW's CGM (e.g., \citealt{Sembach03, Shull09}) and external galaxies (e.g., \citealt{Werk14}) indicate that multiple discrete absorbers exist along sightlines. Our synthetic observations from external views and from within the simulated galaxy also suggest multiple absorbers at discrete velocities. Although the other method that \cite{Werk14} applied takes into account the volume filling factor, this mass estimate comes with large uncertainty since it is largely dependent on the calculation of $\Nhi$ and $\nhi$. 

Thus far, we cannot confirm that the MW is an outlier in CGM mass. Other potential factors, such as the completeness of observations and the gas density profile of the CGM, need to be more fully considered.

\subsubsection{CGM Column Density}
\label{sec5.2.2}
Column densities of metal lines detected in the MW's CGM are generally lower than those of external galaxies. For example, \cite{Tumlinson11} found a typical value of ${\rm log\ N_{OVI}}$=14.5 for 27 out of 30 star-forming galaxies with sSFR$\geq$10$^{-11}$ yr$^{-1}$, which is higher than the mean value of ${\rm log\ \big \langle N_{OVI}\big \rangle}$=14.0 (at sSFR$\sim$2-6$\times$10$^{-11}$ yr$^{-1}$) for the MW's CGM \citep{Sembach03}. To evaluate this discrepancy in column density, variations of path length from observations of the MW's CGM to those of external galaxies need to be considered as well as the velocity obscuration effect.

As an approximation, we compute the column densities of warm-hot gas ${\rm N_{wh}}$ in the simulated halo. We calculate the ratio of ${\rm N_{wh}^{ex}/N_{wh}^{in}}$, where ``ex" indicates column densities of the warm-hot CGM as observed from external views (edge-on, 45$^{\degree}$ and face-on), and ``in" for that with an internal view as described in Section \ref{sec3}. For ${\rm N_{wh}^{in}}$, we sum up all the material along the sightlines and do not separate high velocity and low velocity halo gas so that the influence of path length difference independent of disk obscuration effects can be evaluated. We find that the ratio of external to internal column density varies from 0.9 to 5.0 with a mean of 2.34. This implies that the column densities of CGM in external galaxies are intrinsically higher by a factor of $\sim$2.0 than those of the MW because of the difference in path length. Taking this into account, we can correct the column densities of the MW's CGM to ${\rm log\ \big \langle N_{OVI}\big \rangle}$=14.30, similar to other L$\sim$L$_*$ galaxies. This is a lower limit given that the low velocity gas obscured by MW's disk could make an additional contribution to the average column density. Thus we find the MW's halo detected in \ion{O}{6} is not an outlier in column density. 

\subsection{Obscured Low Velocity {\rm \ion{H}{1}} complexes}
\label{sec5.3}

In this section we examine the obscured halo gas in the context of complexes of HI gas that may exist in the Milky Way halo. These complexes would be low velocity halo clouds (LVHCs), or halo clouds that have LSR velocities indistinguishable from those in the MW disk, but are otherwise akin to HVCs \citep{Wakker91d, Peek08}. These LVHCs may represent bulk flows within the MW halo that have been missed by standard HVC studies due to our location in the disk.  

There is some evidence for MW LVHCs from far-infrared (FIR) studies of low velocity clouds that find a small number have undetectable FIR emission, similar to HVCs \citep{Peek09, Saul14}.  This lack of FIR emission in HVCs is likely due to some combination of low dust content and weak dust heating from the interstellar radiation field. \cite{Peek09} examined the ratio of dust FIR emission to \ion{H}{1} gas column density and found several potential LVHCs, in particular L2 and L5 noted in Figure \ref{fig:f9}. \cite{Saul14} examined this ratio for a larger population of low velocity clouds and identified an entire complex of potential LHVCs in a region of sky near a HVC complex known as the WA complex.   \cite{Saul12} referred to this potential LVHC population as the +WQ3 clouds. As can be seen in the left panel of Figure \ref{fig:f9}, there is a position-velocity link between the +WQ3 clouds, the WA complex, and L2 and L5, indicating these clouds represent a bulk flow that was not previously recognized in halo gas studies.

To examine if we see similar LVHC features in the simulated CGM, we plot the LSR velocities of gas in the simulated CGM for the same patch of sky as the clouds discussed above in the right panel of Figure \ref{fig:f9}. We find a velocity gradient in the simulated galactic sky, which is consistent with the observed pattern from complex WA to the L2, L5 and +WQ3 LVHCs. This implies that those regions with low $\vlsr$ towards the galactic center and anti-galactic center in the simulated halo have populations of LVHCs, similar to what the MW FIR studies seem to be finding. Since the low $\vlsr$ regions (i.e., the low $f_m$ regions in Figure \ref{fig:f5}) are spatially coherent, we expect there are additional large LVHC complexes in the MW's CGM which have been missed entirely due to the velocity cutoffs provided by the disk emission. 

\section{Conclusions}
\label{sec6}
When we observe the MW's CGM, the low velocity gas is missing and we are biased towards high velocity gas due to disk contamination. To assess the amount of obscured low velocity halo gas and its contribution to the MW CGM's baryon mass, we conduct synthetic observations of a simulated MW-mass galaxy. We embed mock observers inside the simulated disk and observe outward to examine gas kinematics as shaped by galactic rotation. We also observe the simulated CGM externally for three different disk inclination angles. The synthetic observations of the simulated CGM with inside-out views and external views show comparable kinematic properties with the MW's CGM and that of other ${\rm L\sim L_*}$ galaxies. 

Our main results can be summarized as follows: 

\begin{enumerate}
\item The mean mass fraction of high velocity (observable) gas in the simulated CGM is 0.55 (Figure \ref{fig:f5}; Table \ref{tb:fm}) as is seen by the mock observer in the simulated disk. This value fluctuates from 0.39 to 0.75 as the location of the mock observer changes.

\item Large spatially-coherent velocity fields are seen in the simulated CGM, where most of the high velocity gas is at low galactic latitudes and the low velocity gas generally exists at higher galactic latitudes. This suggests that large \ion{H}{1} complexes moving at low velocities in the MW's CGM may be missed by observers located at solar radius.

\item Given the similarity between the simulation and the MW's CGM kinematics, we find that the obscuration-corrected mass of the MW's CGM (T$\leq10^6$ K) is 2.8$\times$10$^{8}\Msun$ excluding the Magellanic System. This gives the lower bound of the CGM mass since current mass estimates are limited to the inner halo where reliable distance constraints can be found. Although the corrected mass of the MW's CGM is below the CGM mass of other ${\rm L\sim L_*}$ galaxies, we find similar column densities for \ion{O}{6} lines after correcting for the difference in path length between observations of the MW's CGM and that of external galaxies. 

\item Obscured low velocity gas and observable high velocity gas in the simulated CGM closely overlap in phase space, indicating that this 50$\%$ obscuration in mass affects observations of gas at most phases at ${\rm T<10^6\ K}$ (Figure \ref{fig:f6}). This is to say, observations of the MW's CGM using various metal lines at different ionization states generally miss 50$\%$ of CGM gas because it moves at low velocities. In addition, no significant difference in metallicity is found between the low velocity and high velocity gas. 

\item Breaking the simulated CGM into a series of galactocentric shells, we find that the observable mass fraction does not vary with galactocentric distance (Figure \ref{fig:f6}). The spatially coherent distribution of high velocity gas in the inner and outer halo may significantly reduce the gas observability in the outskirts of the MW's CGM. This happens if distant high velocity halo gas moves at velocities close to that of the nearby halo gas so that their line separations in spectra are too small to recognize.   
\end{enumerate}

We thank Greg Bryan for useful discussions on the interpretation of multiphase gas and kinematics in the simulation throughout this project, and Jessica K. Werk and Nicolas Lehner for helpful discussions on gas column densities. MEP acknowledges funding from NSF Awards AST-1008134 and AST-1312888, and the Clare Boothe Luce Program. MRJ acknowledges funding from AST-1312888. JEGP was supported by HST-HF-51295.01A, provided by NASA through a Hubble Fellowship grant from STScI, which is operated by AURA under NASA contract NAS5-26555. 

\bibliography{main}
\end{document}